\documentclass[twocolumn,showpacs,preprintnumbers,prb,amsmath,amssymb]{revtex4}
\usepackage[dvips]{graphicx,color} 
\usepackage{dcolumn}% Align table columns on decimal point
\usepackage{bm}% bold math

\usepackage{subfigure} % Written by Steven Douglas Cochran
\begin{document}

%\preprint{APS/123-QED}

\title{Free-energy landscape of nucleation with an intermediate metastable phase studied using capillarity approximation}% Force line breaks with \\

\author{Masao Iwamatsu}
\email{iwamatsu@ph.ns.tcu.ac.jp}
\affiliation{
Department of Physics,
Tokyo City University,
Setagaya-ku, Tokyo 158-8557, Japan
%This line break forced with \textbackslash\textbackslash
}%

%\author{Charlie Author}
% \homepage{http://www.Second.institution.edu/~Charlie.Author}
%\affiliation{
%Second institution and/or address\\
%This line break forced% with \\
%}%

\date{\today}% It is always \today, today,
             %  but any date may be explicitly specified

\begin{abstract}
Capillarity approximation is used to study the free-energy landscape of nucleation when an intermediate metastable phase exists.  The critical nucleus that corresponds to the saddle point of the free-energy landscape as well as the whole free-energy landscape can be studied using this capillarity approximation, and various scenarios of nucleation and growth can be elucidated. In this study we consider a model in which a stable solid phase nucleates within a metastable vapor phase when an intermediate metastable liquid phase exists.  We predict that a composite critical nucleus that consists of a solid core and a liquid wetting layer as well as pure liquid and pure solid critical nuclei can exist depending not only on the supersaturation of the liquid phase relative to that of the vapor phase but also on the wetting behavior of the liquid surrounding the solid. The existence of liquid critical nucleus indicates that the phase transformation from metastable vapor to stable solid occurs via the intermediate metastable liquid phase, which is quite similar to the scenario of nucleation observed in proteins and colloidal systems.  By studying the minimum-free-energy path on the free-energy landscape, we can study the evolution of the composition of solid and liquid within nuclei not limited to the critical nucleus.  
\end{abstract}

% PACS, the Physics and Astronomy
                             % Classification Scheme.
%\keywords{Suggested keywords}%Use showkeys class option if keyword
                              %display desired
\maketitle

\section{\label{sec:sec1}Introduction}
A phase transformation involves {\it nucleation} and {\it growth}. When an intermediate metastable phase exists between the initial metastable phase and the final stable phase, the scenario of phase transformation becomes complex and has attracted much interest from more than a decade ago~\cite{Ostwald1897}.  Recently, renewed interest has emerged not only in the field of traditional metallurgy~\cite{Chung2009} but also in the field of soft-condensed-matter physics of proteins and colloids~\cite{Poon2002,Vekilov2004,Sear2007}.  Even for the simple Lennard-Jones system below the triple point, it has recently been suggested ~\cite{van Meel2008} that the intermediate metastable liquid plays a crucial role in the vapor to solid phase transformation.

 The direct microscopic computer simulation of a phase transformation using molecular dynamics or the Monte Carlo method is possible~\cite{van Meel2008,TenWolde1997} but is still a difficult task. To avoid the requirement of huge computational resources and to obtain a qualitative (coarse-grained) picture of the kinetics of a phase transformation, a mesoscopic approach called the phase-field model, which is based on density functional theory~\cite{Talanquer1998,Granasy2000,Sear2001}, has been frequently used. Using this phase-field model for the nonconserved order parameter, Bechhoefer et al.~\cite{Bechhoefer1991} and Celestini and ten Bosch~\cite{Celestini1994} discovered the formation of a finite layer of an intermediate metastable phase at the growing front of the stable phase.  Their results were also confirmed for a different class of free-energy landscape~\cite{Granasy2000}. Later, these works were further extended to the conserved order parameter~\cite{Evans1997a}.  These previous works, however, focused on growth~\cite{Bechhoefer1991,Celestini1994,Evans1997a} rather than nucleation.  Usually the existence of a critical nucleus has been assumed from the outset of the simulation~\cite{Iwamatsu2010}.  

\begin{figure}[htbp]
%Fig.1
\begin{center}
\includegraphics[width=0.7\linewidth]{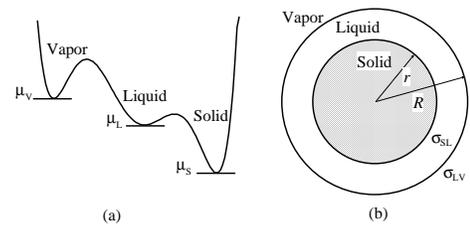}
\caption{
(a) Schematic of the free energy $G$ of a three-phase system along a fictitious reaction coordinate.  (b) Composite nucleus consisting of a stable solid core surrounded by an intermediate metastable liquid layer in a metastable vapor phase. }
\label{fig:1}
\end{center}
\end{figure}

It has been customary to assume two-step nucleation when there is an intermediate metastable phase~\cite{Kashchiev1998,Kashchiev2005,Valencia2004,Valencia2006,Nicolis2003}.  Imagine that nucleation proceeds via two successive processes: the nucleation of a metastable liquid nucleus within a vapor phase, and the subsequent nucleation of a solid phase within the metastable liquid nucleus. Figure~\ref{fig:1}(a) shows a schematic diagram of the free energy $G$ of nucleation along a fictitious one-dimensional reaction coordinate when an intermediate metastable phase exists where a metastable vapor phase (V) transforms into a stable solid phase (S) via an intermediate metastable liquid phase (L).  This diagram implicitly assumes that the reaction coordinate is one-dimensional.  Also, the nucleus of the stable S phase is assumed to grow within the nucleus of the metastable L phase (Fig.~\ref{fig:1}(b)), implying successive nucleation and growth.   

Kashchiev and coworkers~\cite{Kashchiev1998,Kashchiev2005} have assumed the above scenario and considered both nucleation and growth on the same footing within the framework of Kolmogorov-Johnson-Mehl-Avrami (KJMA) kinetics~\cite{Kolmogorov1937, Johnson1939,Avrami1939}.  They assumed that these two nucleation processes occur successively (Fig.~\ref{fig:1}(b)) and derived a formula for the time dependence of the volume fraction of the stable phase.  They found, in particular, that the nucleation of the stable phase can be delayed by the existence of an intermediate metastable phase.  Their result also suggests the appearance of the long-lived metastable intermediate phase predicted by the phase-field model~\cite{Bechhoefer1991,Celestini1994,Granasy2000,Iwamatsu2010}.  Similarly, Valencia and Lipowsky~\cite{Valencia2004,Valencia2006} derived a formula for the nucleation rate assuming the intermediate metastable phase using the double-barrier model by extending the theory of Kramers on the stochastic process of nucleation~\cite{Kramers1940,Gardiner2004}.  Their formula also suggests that the nucleation rate of the stable phase is lower if the intermediate metastable phase exists. On the other hand, Nicolis and Nicolis~\cite{Nicolis2003} showed that the nucleation rate of the stable phase can be enhanced by the existence of the metastable phase by using the theory of Kramers~\cite{Kramers1940,Gardiner2004}. A similar enhancement of the nucleation rate by the existence of an intermediate metastable phase was directly observed by a numerical experiment by Sear~\cite{Sear2009} based on the lattice model . In these theoretical studies~\cite{Kashchiev1998,Kashchiev2005,Valencia2004,Valencia2006,Nicolis2003}, however, it was implicitly assumed that the nucleation reaction coordinate is one-dimensional (Fig.~\ref{fig:1}(a)) and that there are two distinct nucleation barriers (a double barrier).  

In fact, nucleation proceeds on the free-energy surface in multidimensional phase space, and the nucleation barrier corresponds to the saddle point on the multidimensional surface~\cite{Iwamatsu2009,Harrowell2010}.  It is, therefore, not apparent if two distinct saddle points really exist in the free-energy landscape as assumed by previous authors~\cite{Kashchiev1998,Kashchiev2005,Valencia2004,Valencia2006,Nicolis2003}.  Incidentally, our problem resembles that of the vapor phase nucleation of a binary droplet and, therefore, resembles that of the deliquescence~\cite{Djikarev01,Shchekin08,McGraw09}.  In our problem, however, two components, the liquid and the solid, come from the same metastable vapor phase.  On the other hand, the solute and the solvent in the deliquescence come separately from the pre-existing solid core and vapor phase respectively.   

In this study, we use the classical capillarity approximation based on the assumption of uniform density to reduce the multidimensional free-energy surface to a surface in a finite parameter space and study the nucleation pathway when an intermediate metastable phase is involved.  We use a crude capillarity approximation that is also the basis of classical nucleation theory (CNT) as it allows us to study the critical nucleus as well as the whole process of nucleation of a single nucleus.

\section{\label{sec:sec2}Capillarity theory of nucleation with an intermediate metastable phase}

We first consider a composite nucleus consisting of a solid core with radius $r$ embedded in a liquid nucleus of radius $R$ (Fig.~\ref{fig:1}(b)).  This problem is akin to surface melting or wetting~\cite{Broughton1983,Reiss1988,Beaglehole1991,Kofman1993,deGennes1985,Dietrich1988}, where the thickness of the wetting layer is given by $R-r$. In fact, our problem is inverse melting, where a stable substrate appears {\it after} the appearance of the wetting layer.

Within the capillarity approximation, the free energy required to form a liquid nucleus with radius $R$ from a metastable vapor phase is given by
\begin{equation}
\Delta G_{\rm LV}=N\left(\mu_{\rm L}-\mu_{\rm V}\right)+4\pi R^{2}\sigma_{\rm LV},
\label{eq:1-1x}
\end{equation}
where $R$ is the radius of the liquid nucleus, $N$ is the number of liquid molecules within radius $R$, $\sigma_{\rm LV}$ is the liquid-vapor surface tension, and $\mu_{\rm L}$ and $\mu_{\rm V}$ are the chemical potentials of the liquid and vapor phases, respectively. Similarly, the free energy required to form a solid nucleus of radius $r$ within a metastable liquid nucleus of radius $R$ (Fig.~\ref{fig:1}(b)) is given by
\begin{equation}
\Delta G_{\rm SL} = \left(N-N_{1}\right)\left(\mu_{\rm S}-\mu_{\rm L}\right)+4\pi r^{2}\sigma_{\rm SL},
\label{eq:1-2x}
\end{equation}
where $N-N_{1}$ is the number of solid molecules, $N_{1}$ is the number of liquid molecules in the liquid layer with width $R-r$, $\mu_{\rm S}$ is the chemical potential of the stable solid phase, and $\sigma_{\rm SL}$ is the solid-liquid surface tension.

The total free energy $\Delta G$ required to form a composite nucleus with a solid core of radius $r$ surrounded by a liquid layer of width $R-r$ is given by
\begin{equation}
\Delta G = \Delta G_{\rm LV} + \Delta G_{\rm SL} + \Delta G_{\rm SLV},
\label{eq:1-4x}
\end{equation}
where
\begin{equation}
\Delta G_{\rm SLV} = 4\pi R^{2}S e^{-(R-r)/\xi}
\label{eq:1-5x}
\end{equation}
is the correction term due to the short-range interaction between the liquid-vapor and solid-liquid interfaces~\cite{Broughton1983,Beaglehole1991,Kofman1993} which corresponds to the disjoining pressure in the theory of deliqescence~\cite{Shchekin08,McGraw09}, and
\begin{equation}
S = \sigma_{\rm SV}-\sigma_{\rm LV}-\sigma_{\rm SL}
\label{eq:1-6x}
\end{equation}
is the spreading parameter~\cite{deGennes1985}, where $\sigma_{\rm SV}$ is the solid-vapor surface tension, and $\xi$ specifies the range of interaction~\cite{Broughton1983,Beaglehole1991,Kofman1993}. We have used the crudest possible approximation to the short-range interactions for wetting in Eq.~(\ref{eq:1-5x}) since we are most interested in the global picture of nucleation, although a more refined theory for short-range wetting is available.~\cite{Parry2006} 

From Young's equation
\begin{equation}
\sigma_{\rm SV} = \sigma_{\rm SL} + \sigma_{\rm LV}\cos\theta,
\label{eq:1-7x}
\end{equation}
where $\theta$ is the contact angle~\cite{deGennes1985,Dietrich1988}, we have
\begin{equation}
S=\sigma_{\rm LV}\left(\cos\theta-1\right).
\label{eq:1-8x}
\end{equation}
Therefore, the complete wetting of the solid by the liquid is realized when $S\geq 0$, otherwise the solid is incompletely wet by the liquid layer.  Even though Eq.~(\ref{eq:1-8x}) predicts that the complete wetting with $\theta=0$ implies $S=0$, it is well recognized that the surface free energy $\sigma_{\rm SV}$, $\sigma_{\rm SL}$ and $\sigma_{\rm LV}$ will change~\cite{Bonn01} even in the complete wetting regime with $\theta=0$ such that $S$ becomes positive ($S>0$).  Physically, Eq.~(\ref{eq:1-5x}) represents the interaction between the solid-liquid and liquid-vapor interfaces, which is repulsive ($S>0$) when the complete wetting condition is satisfied as these two interfaces repel each other so that the liquid wetting layer intervenes between the solid and vapor phases.  The condition $S>0$ is also known as the condition of surface melting~\cite{Pluis87}.

From Eqs.~(\ref{eq:1-1x}) to (\ref{eq:1-5x}), $\Delta G$ in Eq.~(\ref{eq:1-4x}) becomes
\begin{eqnarray}
\Delta G &=& N\left(\mu_{\rm L}-\mu_{\rm V}\right)+\left(N-N_{1}\right)\left(\mu_{\rm S}-\mu_{\rm L}\right) \nonumber \\
&+&4\pi R^{2}\left(\sigma_{\rm LV}+\sigma_{\rm SL}\left(r^{2}/R^{2}\right)+S e^{-(R-r)/\xi}\right),
\label{eq:1-9x}
\end{eqnarray}
which leads to the free energy of a solid nucleus directly nucleated from the vapor when $r=R$ and $N_{1}=0$:
\begin{equation}
\Delta G = \Delta G_{\rm SV} = N\left(\mu_{\rm S}-\mu_{\rm V}\right)
+4\pi R^{2} \sigma_{\rm SV}.
\label{eq:1-10x}
\end{equation}
By using $N=4\pi R^{3}/3v_{\rm m}$ and maximizing this free energy by solving $\partial \Delta G_{\rm SV}/\partial R=0$, we obtain the free-energy barrier $\Delta G_{\rm SV}^{*}$ and the critical radius $R_{\rm SV}^{*}$ of the solid critical nucleus:
\begin{equation}
\Delta G_{\rm SV}^{*}=\frac{16\pi}{3}\frac{v_{\rm m}^{2}\sigma_{\rm SV}^{3}}{\Delta \mu_{\rm SV}^{2}},
\;\;\;
R_{\rm SV}^{*}=\frac{2v_{\rm m}\sigma_{\rm SV}}{\Delta\mu_{\rm SV}},
\label{eq:1-11x}
\end{equation} 
where $v_{\rm m}$ is the molecular volume and $\Delta\mu_{SV}=\mu_{\rm V}-\mu_{\rm S}>0$.  Similarly, the free energy of a liquid nucleus is obtained when $r=0$ and $N_{1}=N$:
\begin{eqnarray}
\Delta G &=& =\Delta G_{\rm LV}=N\left(\mu_{\rm L}-\mu_{\rm V}\right)+4\pi R^{2}\left(\sigma_{\rm LV}+S e^{-R/\xi}\right) \nonumber \\
&\simeq & N\left(\mu_{\rm L}-\mu_{\rm V}\right)+4\pi R^{2}\sigma_{\rm LV},
\label{eq:1-12x} 
\end{eqnarray}
where the approximation of the second line is valid provided $\xi \ll R$.  The free-energy barrier $\Delta G_{\rm LV}^{*}$ and critical radius $R_{\rm LV}^{*}$ are given by formulae similar to those in Eq.~(\ref{eq:1-11x}) obtained by changing the suffix from "SV" to "LV" and using the chemical potential difference $\Delta\mu_{\rm LV}=\mu_{\rm V}-\mu_{\rm L}$.  The molecular volume $v_{\rm m}$ is assumed to be the same in the solid and liquid nuclei.

However, instead of using the above expressions for the free energy, we introduce the following scaled free energies:
\begin{eqnarray}
g&=&\Delta G/\Delta G_{\rm SV}^{*},\;\;\; g_{\rm LV}=\Delta G_{\rm LV}/\Delta G_{\rm SV}^{*},
\nonumber \\
g_{\rm SL}&=&\Delta G_{\rm SL}/\Delta G_{\rm SV}^{*},\;\;\;g_{\rm SV}=\Delta G_{\rm SV}/\Delta G_{\rm SV}^{*},
\label{eq:1-13x} \\
g_{\rm SLV}&=&\Delta G_{\rm SLV}/\Delta G_{\rm SV}^{*}, \nonumber
\end{eqnarray}
and express these energies in terms of two parameters: the scaled radius $x$ of the liquid nucleus, defined by
\begin{equation}
x = R/R_{\rm SV}^{*},
\label{eq:1-13y}
\end{equation}
and the proportion $t$ of the solid radius $r$ relative to the liquid radius $R$,
\begin{equation}
t = r/R. 
\label{eq:1-13z}
\end{equation}
By varying $t$ in the range $0\leq t \leq 1$, we can study a composite nucleus composed of a solid core surrounded by a liquid wetting layer.  The nucleus is all solid when $t=1$ and all liquid when $t=0$.

The above free energies as functions of the two parameters $(x,t)$ are given by
\begin{eqnarray}
g\left(x,t\right)&=&-2\delta x^{3}-2\left(1-\delta\right)\left(xt\right)^{3},
\label{eq:1-14x} \\
&+&3x^{2}\left(\beta +\alpha t^2 + \left(1-\left(\alpha+\beta\right)\right)e^{-x(1-t)/\tau}\right)
\nonumber
\end{eqnarray}
and
\begin{eqnarray}
g_{\rm SV} &=& g\left(x,1\right)=-2x^3+3x^2, \nonumber \\
g_{\rm LV} &=& g\left(x,0\right)=-2\delta x^{3} + 3x^{2}\left(\beta+\left(1-\left(\alpha+\beta\right)\right)e^{-x/\tau}\right), 
\nonumber \\
g_{\rm SL}&=& -2\left(1-\delta\right)\left(xt\right)^{3}+3\alpha \left(xt\right)^{2},
\label{eq:1-15x} \\
g_{\rm SLV}&=&3x^{2}\left(1-\left(\alpha+\beta\right)\right)\left(e^{-x(1-t)/\tau}-e^{-x/\tau}\right),
\nonumber
\end{eqnarray}
where we have included the exponential correction of $\Delta G_{\rm SLV}$ (Eq.~(\ref{eq:1-5x})) in the definition of $g_{\rm LV}$ (the last term of Eq.~(\ref{eq:1-12x})), and introduced the material parameters
\begin{eqnarray}
\alpha&=&\sigma_{\rm SL}/\sigma_{\rm SV},\;\;\;\beta=\sigma_{\rm LV}/\sigma_{\rm SV},
\nonumber \\
\delta&=&\Delta \mu_{\rm LV}/\Delta \mu_{\rm SV},\;\;\; \tau=\xi/R_{\rm SV}^{*}.
\label{eq:1-16x}
\end{eqnarray}

The total free energy of the composite nucleus [Eq.~(\ref{eq:1-14x})] is rewritten as
\begin{equation}
g = g_{\rm LV} + g_{\rm SL} + g_{\rm SLV}.
\label{eq:1-16xx}
\end{equation}
The spreading parameter $S$ is given by
\begin{equation}
S/\sigma_{\rm SV}=1-\left(\alpha+\beta\right),
\label{eq:1-17x}
\end{equation}
using the reduced surface tensions $\alpha$ and $\beta$. From the Dupr\'e equation~\cite{Israelachvili1992}, $\sigma_{\rm SL} \approx \sigma_{\rm SV}+\sigma_{\rm LV} - \sqrt{\sigma_{\rm SV}\sigma_{\rm LV}}=\left(\sqrt{\sigma_{\rm SV}}-\sqrt{\sigma_{\rm LV}}\right)^2$, we have $\alpha\approx \left(1-\sqrt{\beta}\right)^2$.  Since the density of the solid phase is close to that of the liquid phase except near the critical point, we expect that $\sigma_{\rm SV}\approx \sigma_{\rm LV}$, $\beta \approx 1$, and $\alpha\ll 1$.  Incomplete wetting ($S<0$) is realized when $\alpha+\beta>1$.

The critical radius $x_{\rm SV}^{*}$ and activation energy $g_{SV}^{*}$ of a solid nucleus that is directly nucleated from vapor are given by Eq.~(\ref{eq:1-11x}) and are written as 
\begin{equation}
x_{\rm SV}^{*}=1,\;\;\; g_{SV}^{*}=1.
\label{eq:1-18x}
\end{equation}
Similarly, formulae for a metastable liquid nucleus in vapor can be obtained from $\partial g_{\rm LV}/\partial x=0$ and are approximately given by
\begin{equation}
x_{\rm LV}^{*}\simeq \frac{\beta}{\delta},\;\;\; g_{\rm LV}^{*}\simeq \frac{\beta^{3}}{\delta^{2}}.
\label{eq:1-19x}
\end{equation}
On the other hand, those for a solid nucleus that is nucleated from a metastable liquid phase can be obtained from $\partial g_{\rm SL}/\partial t=0$ and are given by
\begin{equation}
t_{\rm SL}^{*}=\frac{\alpha}{x(1-\delta)},\;\;\;g_{\rm SL}^{*}=\frac{\alpha^{3}}{\left(1-\delta\right)^{2}}.
\label{eq:1-19y}
\end{equation}
If $\beta^{3}/\delta^{2} > 1$, the initial critical nucleus will be mostly solid as $g_{\rm LV}^{*}>g_{\rm SV}^{*}$.  This scenario is expected when $\beta\approx 1 > \delta$ and the metastable liquid phase is closer to the metastable vapor phase than the stable solid phase ($\delta=\Delta\mu_{\rm LV}/\Delta\mu_{\rm SV}< 1$).  Since the liquid phase is less stable, the solid phase will be directly nucleated from the vapor phase.

On the other hand, when $\beta^{3}/\delta^{2} < 1$ the initial critical nucleus will be liquid.  In this case, the metastable liquid phase is closer to the stable solid phase ($\delta\approx 1$).  A stable solid nucleus will grow within the metastable liquid matrix, and two-step nucleation~\cite{Kashchiev1998,Valencia2004,Kashchiev2005} with double barriers is expected because the most stable phase is the solid.  

Therefore, by increasing the relative supersaturation $\delta$ of the liquid phase, one may expect the bifurcation from a solidlike critical nucleus surrounded by a thin wetting layer of liquid for small $\delta$ to a liquidlike critical nucleus for large $\delta$.  This solidlike to liquidlike change of the character of the critical nucleus was theoretically predicted by Gr\'an\'asy and Oxtoby~\cite{Granasy2000} using the triple-parabola model, and is implied by Fig. 4 of Talanquer and Oxtoby~\cite{Talanquer1998} using density functional theory. However, they only considered the critical nucleus at the saddle point of the free-energy landscape because density functional theory can only be used to study the critical nucleus that corresponds to the stationary state at the saddle point on the free-energy landscape.  The evolution of the nucleus and its composition other than that of the critical nucleus can only be determined through the examination of the free-energy landscape.  To study the nucleation scenario in the free-energy landscape qualitatively, we apply our capillarity theory in the next section.

\section{\label{sec:sec3}Results and Discussion}

\subsection{Incomplete wetting ($\alpha+\beta>1$)}

\begin{figure}[htbp]
\begin{center}
\subfigure[$\delta=-0.2$]
{
\includegraphics[width=0.65\linewidth]{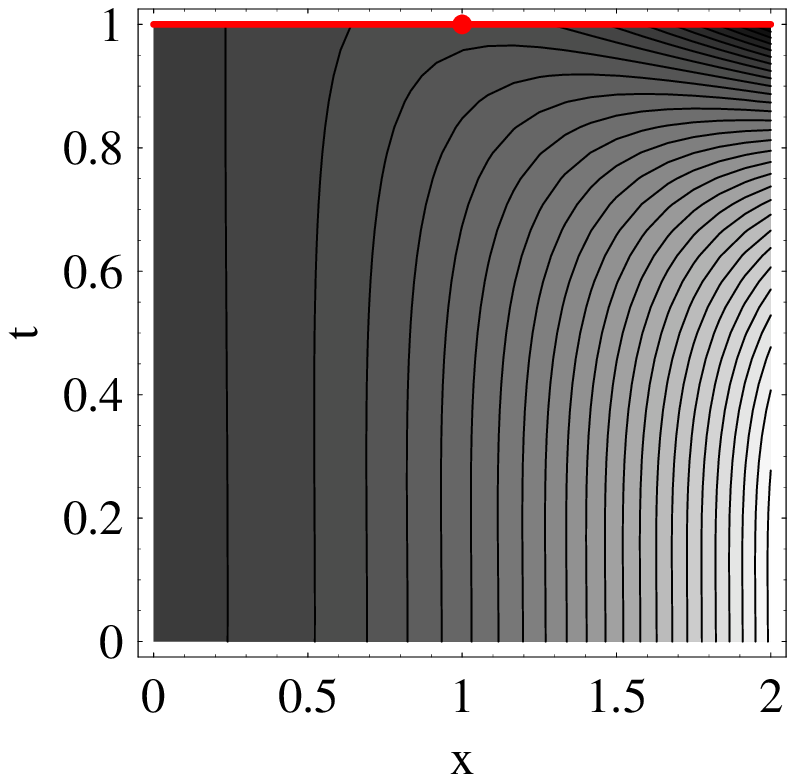}
%\vspace{3cm}
\label{fig:2a}}
\subfigure[$\delta=0.5$]
{
\includegraphics[width=0.65\linewidth]{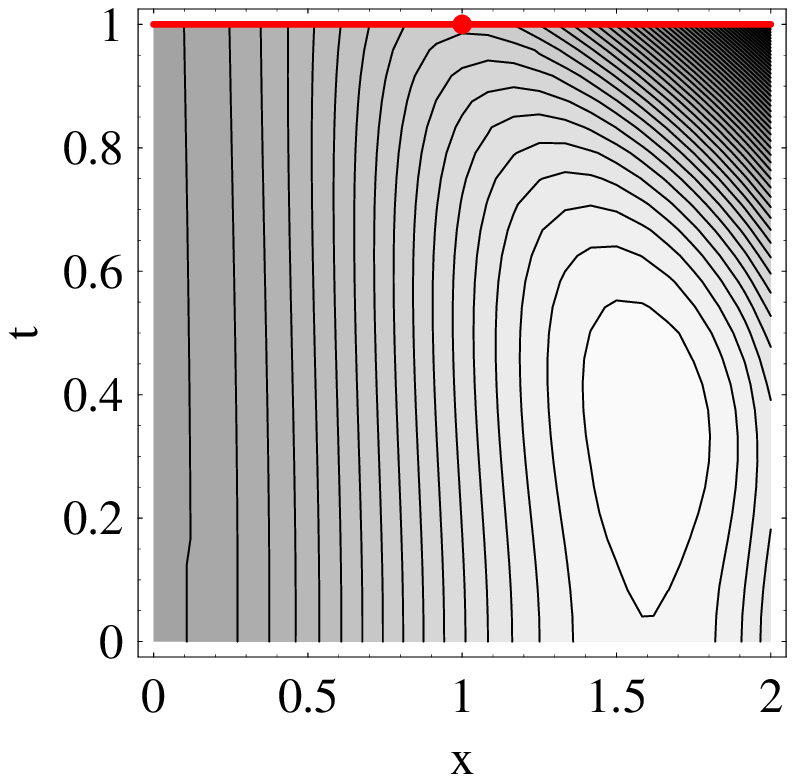}
%\vspace{3cm}
\label{fig:2b}}
\end{center}
\caption{
Contour plots of the free-energy landscape of $g(x,t)$ when the relative supersaturation is low:  (a) $\delta=-0.2$ and (b) $\delta=0.5$. The metastable liquid phase cannot appear and a solid nucleus directly grows within the metastable vapor phase.  The other parameters used are $\alpha=0.3$, $\beta=0.8$, and $\tau=0.2$. The red solid line along $t=0$ (the solid axis) indicates the minimum-free-energy path (MFEP) of the nucleation process. The saddle points are indicated by the red point on each path.  }
\label{fig:2}
\end{figure}

In this case, the metastable liquid may not wet the solid nucleus.  Therefore, it will be unfavorable for the solid phase to nucleate within the liquid phase.  Thus, the composite nucleus is not expected to appear.  Figures \ref{fig:2}(a) and (b) show contour plots of the free-energy landscape of $g(x,t)$ ((a) $\delta=-0.2$, (b) $\delta=0.50$) in the $x-t$ plane when the chemical potential $\mu_{\rm L}$ of the liquid is closer to that ($\mu_{\rm V}$) of the vapor than the solid ($\mu_{\rm S}$).  The nucleation pathway starts at any point along $x=0$ and ends at $x=\infty$ and $t=0$ as the solid phase is the stable phase.  We used the reduced surface tensions $\alpha=0.3$ and $\beta=0.8$, which correspond to the incomplete wetting condition $S>0$. The parameter $\tau$ is fixed to $\tau=0.2$.  
 
Note that the pathway along the $t=0$ axis is the liquid axis, which corresponds to the nucleation of a metastable liquid nucleus, and that along the $t=1$ axis is the solid axis, which corresponds to the direct nucleation of a solid nucleus from the vapor.  A negative $\delta$ means that the liquid phase is less stable than the vapor phase.  In both Fig.~\ref{fig:2}(a) and Fig.~\ref{fig:2}(b) we can locate the saddle point at $x\simeq 1$ and $t=1$, indicated by red points on the minimum-free-energy paths (MFEPs) indicated by the red solid lines. This means that the solid phase nucleates directly from the vapor without passing through the liquid phase.  Since the metastable liquid phase is closer to the vapor phase and is less stable, the metastable liquid phase cannot appear.  Instead, a solid nucleus directly nucleates from the vapor phase as expected from the discussion in the last section.  This solid nucleus does not accompany the liquid wetting layer as the liquid does not wet the solid.

\begin{figure}[htbp]
\begin{center}
\subfigure[$\delta=0.72$]
{
\includegraphics[width=0.65\linewidth]{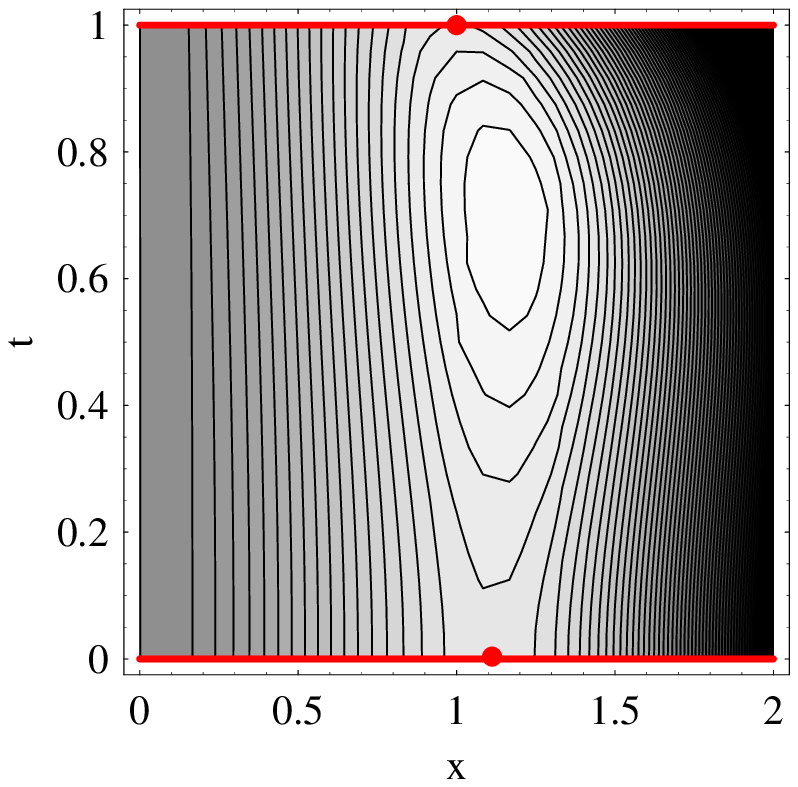}
%\vspace{3cm}
\label{fig:3a}}
\subfigure[$\delta=0.80$]
{
\includegraphics[width=0.65\linewidth]{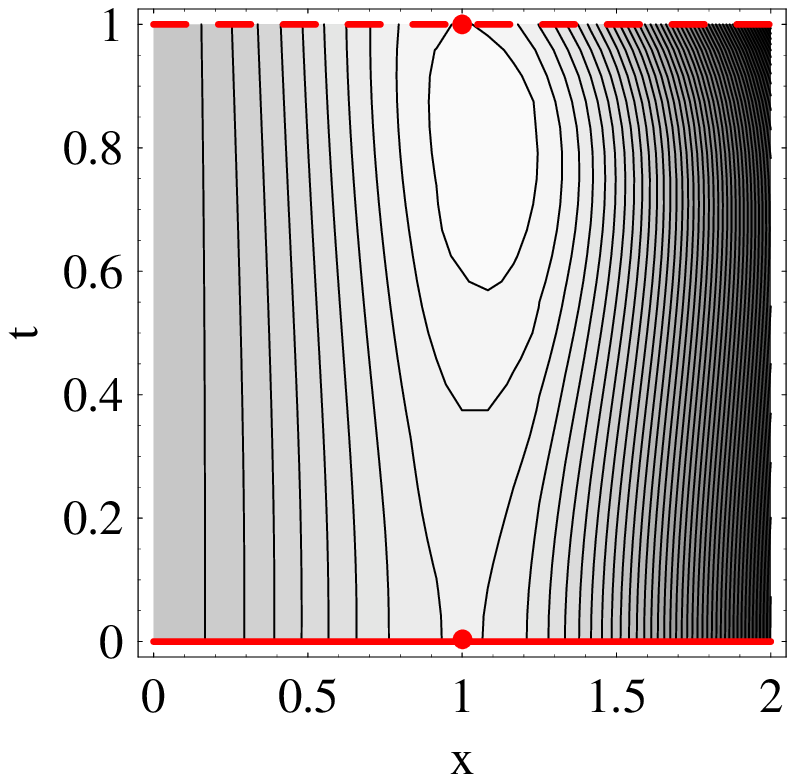}
%\vspace{3cm}
\label{fig:3b}}
\end{center}
\caption{
Contour plots of the free-energy landscape of $g(x,t)$ for the nucleation of the composite nucleus for the moderate and high relative supersaturations: (a) $\delta=0.72$ and (b) $\delta=0.80$ when $\alpha=0.3$, $\beta=0.8$, and $\tau=0.2$. The red solid curve along the solid axis ($t=0$) indicates the MFEP of the nucleation process. The red broken line along the liquid axis ($t=1$) indicates another MFEP.  The saddle points are indicated by the two points on the paths.  }
\label{fig:3}
\end{figure}

Figures \ref{fig:3}(a) and (b) show contour plots of the free-energy landscape when the relative supersaturation $\delta$ of the liquid phase is higher ($\delta=0.72$ and $\delta=0.80$). Parameters $\alpha$, $\beta$, and $\tau$ are the same as those in Fig.~\ref{fig:2}.  In Fig.~\ref{fig:3}(a), we can clearly observe the existence of two saddle points on the solid axis ($t=1$) at $(x,t)=(1.0,1.0)$ and the liquid axis ($t=0$) at $(x,t)=(1.1,0.0)$ indicated by red points on the MFEPs.  The free-energy barrier at these two saddle points is the same ($g^{*}=1.0$).  Therefore, two nucleation pathways (MFEPs), indicated by the two solid lines on the liquid ($t=0$) and solid ($t=1$) axes, coexist.  

Of course, the MFEP along the liquid axis will eventually merge into the MFEP along the solid axis far from the saddle point as the free energy of the solid phase at $(x\rightarrow\infty,t=1)$ should be lower than that of the liquid phase at $(x\rightarrow\infty,t=0)$.  

\begin{figure}[htbp]
\begin{center}
\subfigure[Cross sections along $t=0$ and $t=1$ ($\delta=0.80$)]
{
\includegraphics[width=0.65\linewidth]{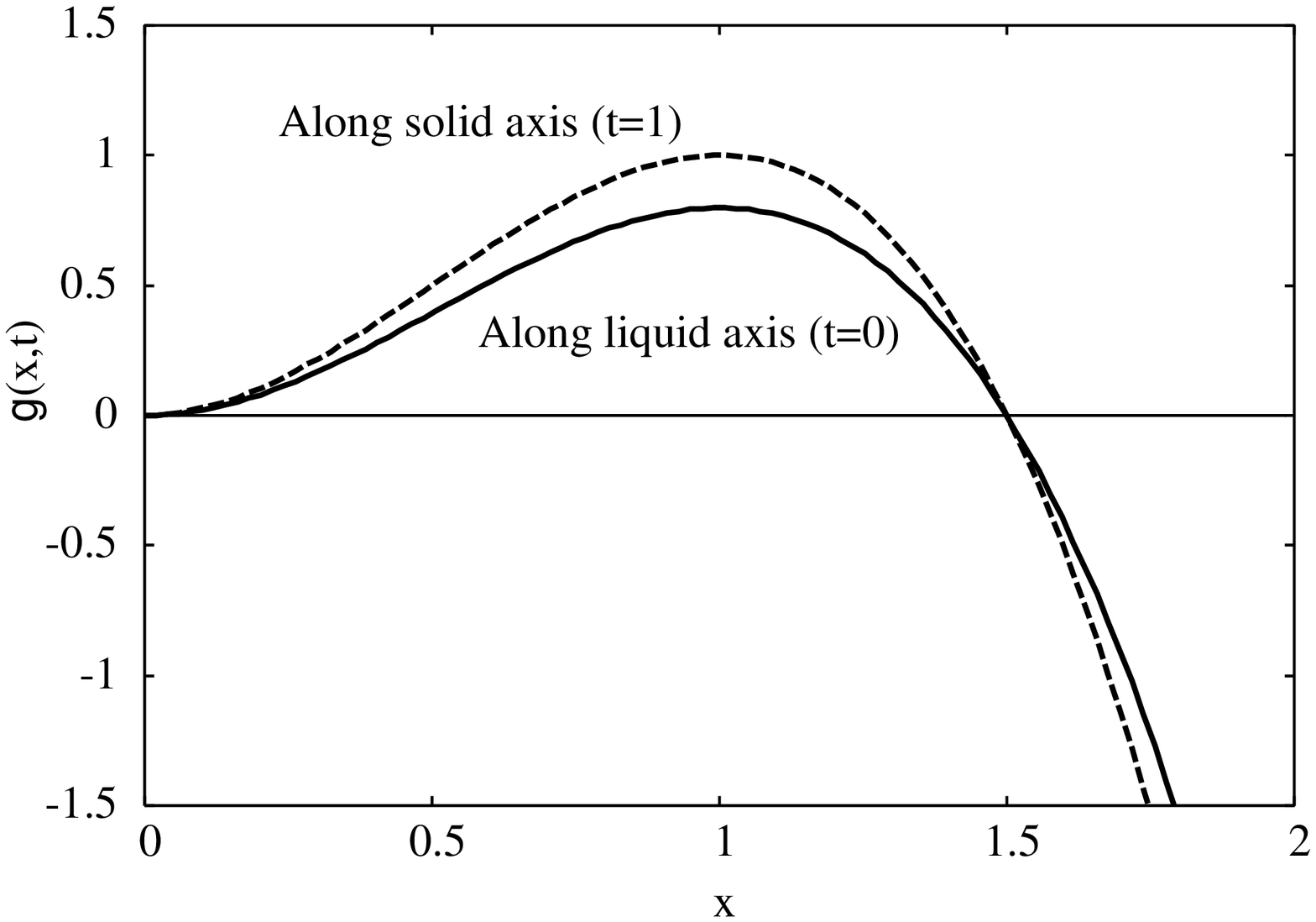}
%\vspace{3cm}
\label{fig:4a}}
\subfigure[MFEP of supercritical nucleus ($\delta=0.80$)]
{
\includegraphics[width=0.65\linewidth]{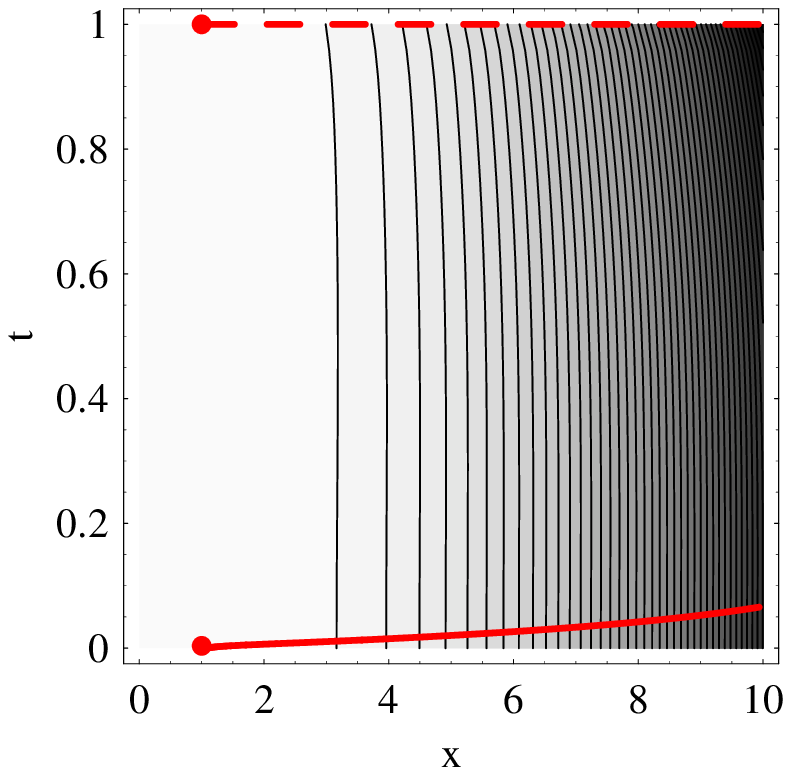}
%\vspace{3cm}
\label{fig:4b}}
\end{center}
\caption{
(a) Cross sections of the free-energy surface along the solid axis ($t=1$) and liquid axis ($t=0$) when the relative supersaturation is high ($\delta=0.80$).  Clearly the saddle point (maximum) of the free-energy curve along the liquid axis is lower than that along the solid axis. (b) Contour plot of the free-energy landscape of $g(x,t)$ and the MFEP (red solid curve) for the supercritical nucleus. The MFEP eventually converges to the upper right corner of the pure solid phase.   }
\label{fig:4}
\end{figure}

As the relative supersaturation $\delta$ is further increased and the metastable liquid phase becomes more stable, the saddle point at $(x,t)=(1.0,0.0)$ on the liquid axis becomes lower ($g^{*}=0.80$) than that ($g^{*}=1.0$) at $(x,t)=(1.0,1.0)$ on the solid axis (Fig.~\ref{fig:3}(b)).  Figure \ref{fig:4}(a) shows cross sections of the free-energy surface along the solid axis ($t=1$) and liquid axis ($t=0$) when $\delta=0.72$.  When $\delta$ becomes larger and the liquid phase becomes as stable as the solid phase, the liquid nucleus along the red solid line on the liquid axis becomes more probable than that along the red broken line on the solid axis (Fig.~\ref{fig:3}(b)).  Again, the MFEP along the liquid axis will eventually turn toward that along the stable solid phase at ($x\rightarrow\infty$, $t=0$) far from the saddle point as shown in Fig.~\ref{fig:4}(b).  Therefore, the nucleation pathway from metastable vapor to stable solid through the intermediate metastable liquid state is free-energetically easier than that from vapor to solid directly. Similar results that the nucleation pathway with intermediate state has a lower free energy are obtained by ten Wolde and Frenkel~\cite{TenWolde1997} using Monte Carlo simulation and by Lutsko and Nicolis~\cite{Lutsko2006} using density functional theory.

There are now two pathways of nucleation.  One is the direct nucleation of the solid phase from the metastable vapor.  The other is the indirect nucleation of the metastable liquid phase from the vapor, which will eventually transform into the solid phase without crossing the barrier as shown in Fig.~\ref{fig:4}(b).  Therefore, there are two nucleation rates.  One is the vapor to liquid nucleation rate,
\begin{equation}
J_{\rm LV} = A_{\rm LV} \exp\left(-\frac{\Delta G_{\rm LV}^{*}}{kT}\right),
\label{eq:2-1x}
\end{equation}  
which expresses the nucleation rate of a liquid nucleus, where $\Delta G_{\rm LV}^{*}$ is the free-energy barrier for the liquid nucleus, given by Eq.~(\ref{eq:1-19x}), and $kT$ is the temperature.  $A_{\rm LV}$ is the so-called preexponential factor which is the kinetic factor of molecules attaching to and detaching from a nucleus. The other is the vapor to solid nucleation rate, given by
\begin{equation}
J_{\rm SV} = A_{\rm SV} \exp\left(-\frac{\Delta G_{\rm SV}^{*}}{kT}\right),
\label{eq:2-2x}
\end{equation} 
where  $\Delta G_{\rm SV}^{*}$ and $A_{\rm SV}$ are the free-energy barrier and the preexponential factor similar to those in Eq.~(\ref{eq:2-1x}).

Previously, it has been assumed that nucleation occurs on a one-dimensional energy landscape through the successive nucleation of the liquid from vapor with nucleation rate $J_{\rm LV}$ followed by liquid to solid nucleation with rate $J_{\rm SL}$, given by
\begin{equation}
J_{\rm SL} = A_{\rm SL}\exp\left(-\frac{\Delta G_{\rm SL}^{*}}{kT}\right).
\label{eq:2-3x}
\end{equation} 
where  $\Delta G_{\rm SL}^{*}$ and $A_{\rm SL}$ are the free-energy barrier and the preexponential factor similar to those in Eq.~(\ref{eq:2-1x}).   Using the kinetic theory of nucleation, where the total balance of attachment and detachment of molecules is considered~\cite{Wu1997}, a formula similar to the conductance of two registers connected in series applies approximately for the total nucleation rate $J$~\cite{Valencia2006}:
\begin{equation}
\frac{1}{J} \approx \frac{1}{J_{\rm LV}}+\frac{1}{J_{\rm SL}}.
\label{eq:2-3xx}
\end{equation}
where only the contribution around the two saddle points for  $J_{\rm SL}$ and $J_{\rm LV}$ are retained~\cite{Valencia2004,Valencia2006}.  Therefore, the total nucleation rate $J$ is slower (lower) than $J_{\rm LV}$.  However, in our model two nucleation processes occur in parallel.  Also, liquid to solid nucleation occurs after vapor to liquid nucleation without crossing the energy barrier. Therefore, a formula similar to the conductance of two registors connected in parallel follows:
\begin{equation}
J \approx J_{\rm LV} + J_{\rm SV},
\label{eq:2-4x}
\end{equation}
where the interference of two channels shown in Figs.~\ref{fig:3}(b) and \ref{fig:4}(b) as the solid and the broken line is neglected.  Then, the total nucleation rate $J$ is faster (higher) than $J_{\rm LV}$. The nucleation can be enhanced because not only a direct vapor to solid nucleation channel exists but also the vapor to liquid nucleation is followed by a barrierless liquid to solid transformation.  A similar but slightly different explanation of the enhancement of the nucleation rate due to the presence of a metastable phase near the spinodal has been proposed~\cite{Vekilov2004}.

\subsection{Complete wetting ($\alpha+\beta<1$)}

The critical nucleus is either solid or liquid for the incomplete wetting case because a composite nucleus that consists of a solid nucleus surrounded by a metastable liquid wetting layer is energetically unfavorable.  In contrast, such a composite nucleus is expected to occur for the complete wetting case.

\begin{figure}[htbp]
\begin{center}
\subfigure[$\delta=0.3$]
{
\includegraphics[width=0.65\linewidth]{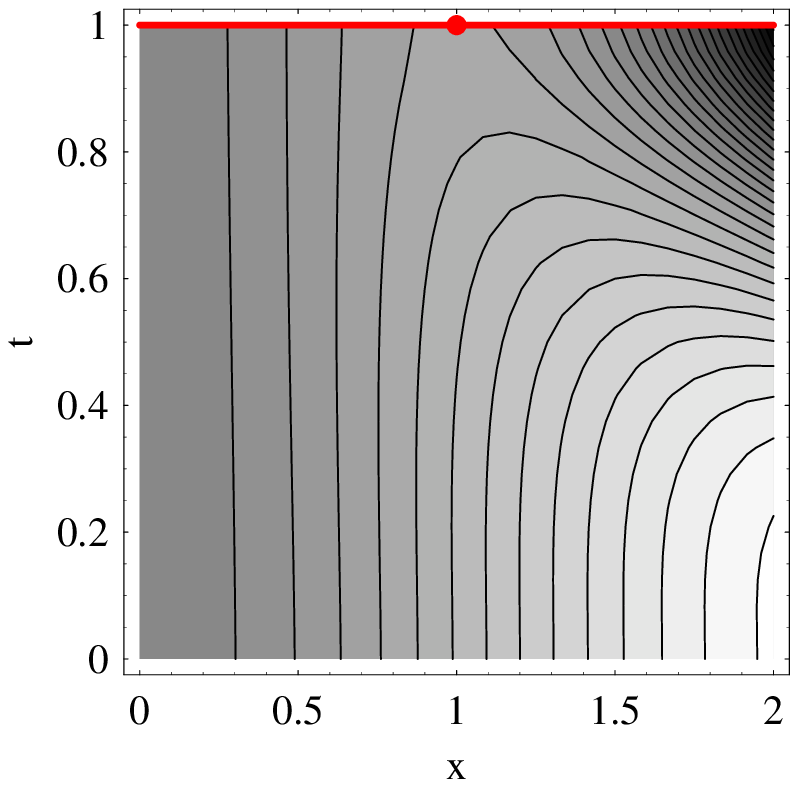}
%\vspace{3cm}
\label{fig:5a}}
\subfigure[$\delta=0.5$]
{
\includegraphics[width=0.65\linewidth]{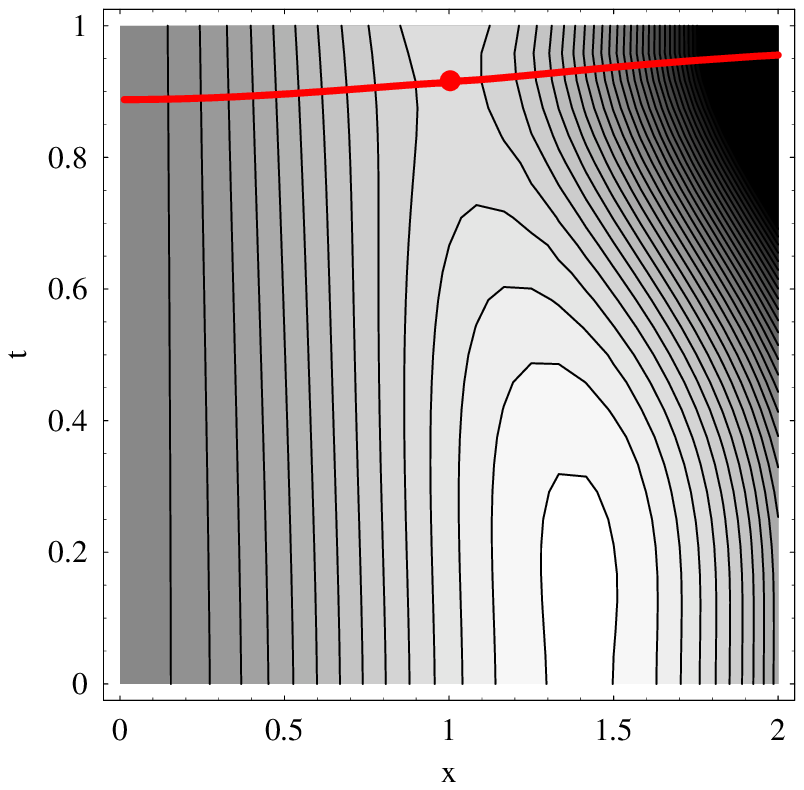}
%\vspace{3cm}
\label{fig:5b}}
\end{center}
\caption{
Contour plots of the free-energy landscape of $g(x,t)$ when the relative supersaturation is low: (a) $\delta=0.3$ and (b) $\delta=0.5$ for the complete wetting case. The critical nucleus at the saddle point is a solid core surrounded by a liquid wetting layer ($\delta=0.5$). The parameters used are $\alpha=0.1$, $\beta=0.7$, and $\tau=0.2$. The red solid lines indicate the MFEP of the nucleation process. The saddle points are indicated by the red point on each path.  }
\label{fig:5}
\end{figure}

Figures \ref{fig:5}(a) and (b) show contour plots of the free-energy landscape of $g(x,t)$ in the $x-t$ plane when the relative supersaturation $\delta$ of the liquid phase is relatively low ((a) $\delta=0.3$ and (b) $\delta=0.5$) when $\alpha=0.1$ and $\beta=0.7$, which satisfy the complete wetting condition ($S/\sigma_{\rm SV}=1-(\alpha+\beta)=1-(0.1+0.7)=0.2>0$).  The red solid line and the red point on each line indicate the MFEP and the saddle point, respectively. When $\delta=0.3$, the saddle point is still on the solid axis.  However, the saddle point shifts from the solid axis to the liquid side at $(x^{*},t^{*})=(1.00,0.92)$ with a free-energy barrier of nucleation $g^{*}=0.98$, which is lower than $g_{\rm SV}^{*}=1.00$ when $\delta=0.5$. These MFEPs (red solid curves) were obtained by solving the overdamped equation of motion of evolution similar to the phase-field equation~\cite{Iwamatsu2009}.  More sophisticated numerical methods such as the string method~\cite{E2007} are unsutable because the basin of attractor that corresponds to the bulk solid phase is located at infinity ($x=\infty$).  

In this case ($\delta=0.5$), the critical nucleus at the saddle point is mostly a solid core surrounded by a thin layer of liquid.  The radius of the solid core occupies 92\% of the total radius of the composite critical nucleus at the saddle point.  After crossing the barrier, the MFEP approaches the bulk solid phase at $(x\rightarrow\infty,t=1)$, and the composite supercritical nucleus becomes a solid nucleus.  Therefore, the nucleation process is not a two-step process~\cite{Kashchiev1998,Valencia2004,Kashchiev2005} but a one-step process with a single activation energy even though a macroscopically thick layer appears around the solid core during the evolution.

\begin{figure}[htbp]
\begin{center}
\includegraphics[width=0.65\linewidth]{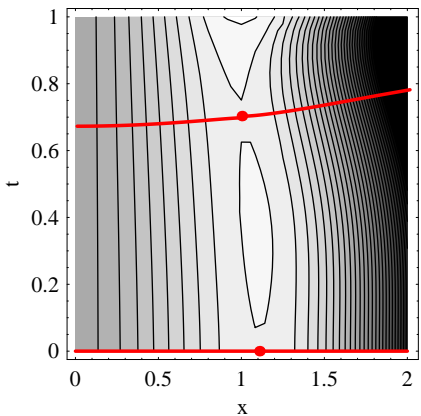}
\end{center}
\caption{
Contour plot of the free-energy landscape of $g(x,t)$ when the relative supersaturation is moderate ($\delta=0.63$) for the complete wetting case. The parameters used are $\alpha=0.1$, $\beta=0.7$, and $\tau=0.2$. The two red solid curves indicate the two transition paths of the nucleation process. The saddle points are indicated by red points on the paths.  There are two critical nuclei. One is a solid core surrounded by a liquid wetting layer and the other is a pure liquid nucleus. 
}
\label{fig:6}
\end{figure}

\begin{figure}[htbp]
\begin{center}
\subfigure[Three-dimensional view of the energy landscape ($\delta=0.63$)]
{
\includegraphics[width=0.75\linewidth]{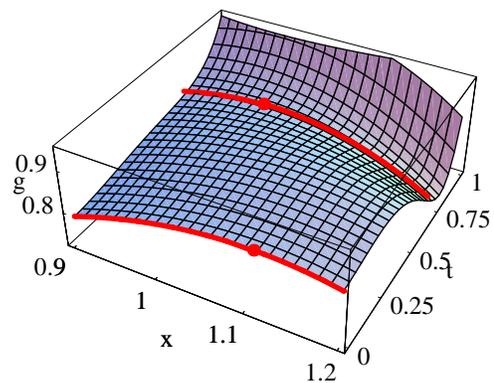}
%\vspace{3cm}
\label{fig:7a}}
\subfigure[Free-energy landscape of supercritical nucleus ($\delta=0.63$)]
{
\includegraphics[width=0.65\linewidth]{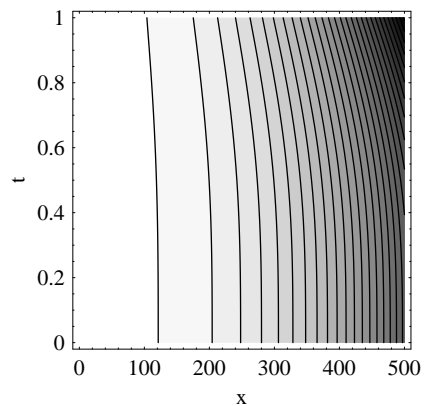}
%\vspace{3cm}
\label{fig:7b}}
\end{center}
\caption{
(a) Three-dimensional view of the free-energy surface shown in Fig.~(\ref{fig:6}) near the saddle point when the relative supersaturation is moderate ($\delta$=0.63).  Clearly there is an energy barrier between the valley along the liquid axis ($t=0$) and the valley near the solid axis ($t\sim 1$).  (b) Contour plot of the free-energy landscape of $g(x,t)$ for the supercritical nucleus.  Two MFEPs eventually converge to the upper right corner of the pure solid phase. }
\label{fig:7}
\end{figure}

As we increase $\delta$ further, two saddle points again appear at $(x,t)=(1.003,0.703)$ and $(x,t)=(1.109,0.000)$ when $\delta=0.63$ (Fig.~\ref{fig:6}). The free-energy barrier at these two points is exactly the same ($f^{*}=0.867$).  Therefore, two nucleation routes coexist. This situation is similar to that in Fig.~\ref{fig:3}(a).  However, since the liquid phase can wet the solid nucleus, one of the critical nuclei is a composite nucleus with a solid core surrounded by a liquid layer whose thickness decreases as the nucleus grows after crossing the barrier.  The other is a pure liquid nucleus.  These two types of nucleus cannot transform between each other freely around the saddle point as there are energy barriers between the two valleys along the liquid axis and near the solid axis as shown in Fig.~\ref{fig:7}(a).

As there is an energy barrier between the two valleys along $t=0$ and $t\sim 1$ near the saddle points, there is a ridge between the two valleys in Figs.~\ref{fig:6} and \ref{fig:7}.  Thus, the nucleation of the stable solid from the pure liquid nucleus at the saddle point along the liquid axis occurs only by overcoming the free-energy barrier from $t=0$ to $t\sim 1$ after crossing the saddle point at $(x,t)=(1.109,0.000)$.  Thus, the nucleation of the stable solid via the liquid critical nucleus appears to occur via two-step nucleation.  In fact, the second barrier is not exactly the saddle point as assumed in simplified theories of two-step nucleation~\cite{Kashchiev1998,Valencia2004,Kashchiev2005}, but is the ridge.  In this case, the liquid supercritical nucleus can survive and become long-lived and macroscopically large even though it is thermodynamically metastable. Of course, the MFEP along this liquid axis should eventually merge into the MFEP along the solid axis far from the two saddle points because the free energy of the solid phase at $(x\rightarrow\infty,t=1)$ should be lower than that of the liquid phase at $(x\rightarrow\infty,t=0)$.

Figure \ref{fig:7}(b) shows a contour plot of the free-energy landscape of the supercritical nucleus far from the saddle point.  One can easily imagine that the MFEP along the liquid axis eventually merges with the MFEP along the solid axis similarly to in Fig.~\ref{fig:4}(b).  In this case, however, we did not calculate the MFEP owing to the numerical difficulty as the length scale is longer than that in Fig.~\ref{fig:4}(b).  Again, the nucleation process is not a two-step process~\cite{Kashchiev1998,Valencia2004,Kashchiev2005} but two parallel channels exist, both of which correspond to one-step nucleation with single activation energies.  Thus, the nucleation rate is given by the formula for the parallel conductance in Eq.~(\ref{eq:2-4x}). 

\begin{figure}[htbp]
%Fig.1
\begin{center}
\includegraphics[width=0.65\linewidth]{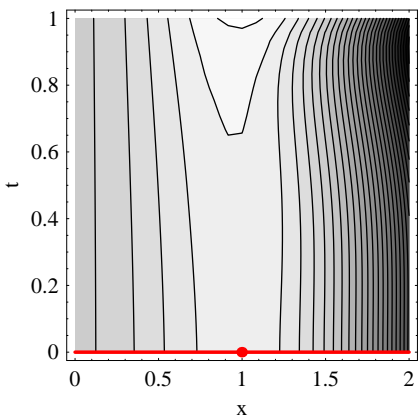}
\caption{
Contour plot of the free-energy landscape of $g(x,t)$ when the relative supersaturation is high ($\delta=0.7$) for the complete wetting case.  The critical nucleus is all liquid along the liquid axis. 
}
\label{fig:8}
\end{center}
\end{figure}

As $\delta$ is further increased and the metastable liquid phase becomes more stable, the saddle point shifts along the liquid axis to $(x,t)=(0.997,0.00)$.  Figure \ref{fig:8} shows a contour plot of the free-energy landscape when $\delta=0.70$.  In this case, the critical nucleus is all liquid.  However, it gradually becomes the solid after crossing the saddle point without overcoming the free-energy barrier.

As the nucleation is expected to proceed via the MFEP~\cite{Iwamatsu2009}, we can summarize the scenario of nucleation as follows: When the relative supersaturation of the liquid phase is low, the liquid phase is metastable and cannot appear, and the solid nucleus or the solid nucleus surrounded by a thin wetting layer of liquid directly appears from the supersaturated vapor (Figs.~\ref{fig:2}(a), (b), and \ref{fig:5}(a), (b)).  In this case, we may expect one-step nucleation of the stable solid phase.  When the supersaturation is moderate, two nucleation channels, along the liquid axis and near the solid axis, coexist (Figs.~\ref{fig:3}(a) and \ref{fig:6}).  When the complete wetting condition is satisfied, the critical nucleus near the solid axis is a composite nucleus that consists of a solid core surrounded by a liquid layer (Figs.~\ref{fig:6} and {\ref{fig:7}(a)}).  In this case, we may expect one-step nucleation again.  However, we may expect that two nucleation processes occur in parallel and that Eq.~(\ref{eq:2-4x}) applies. Therefore, the nucleation rate is enhanced, in general, compared with that of two-step series nucleation given by Eq.~(\ref{eq:2-3x})~\cite{Valencia2004,Valencia2006}. 

When the relative supersaturation of the liquid phase is high, the liquid phase is relatively stable.  Thus, the critical liquid nucleus may first appear within the supersaturated vapor and then pass through the saddle point and continue to grow (Figs.~\ref{fig:3}(b), \ref{fig:4}(a), (b), and \ref{fig:8}).  This growing liquid supercritical nucleus gradually changes to the stable solid phase far from the saddle point without crossing the energy barrier. This scenario of one-step nucleation is completely different, for example, from the nucleation of protein crystals~\cite{Kashchiev1998,Vekilov2004,Kashchiev2005}, where two-step nucleation with two energy barriers is assumed.  

It is well known that the observed nucleation is usually taking place via the metastable intermediate state~\cite{Poon2002,Vekilov2004,Sear2007}.  In fact, our capillarity theory is consistent to this picture because it predicts that the nucleation may occur via the metastable liquid phase when the relative supersaturation of the liquid phase is high (Figs.~\ref{fig:3}(b) and \ref{fig:8}).  Since the metastable liquid phase is free-energetically closer to the stable solid phase than the metastable vapor phase, the nucleation occur from the vapor phase to solid phase via the metastable liquid phase as the liquid is almost stable.  In this case, we may expect one step nucleation from vapor to liquid with a nucleation rate $J_{\rm LV}$ given by a single free energy barrier $\Delta G_{\rm LV}^{*}$.  Even though there are no true saddle point for the liquid to solid transition (Fig.~\ref{fig:4}(b), \ref{fig:7}(b) and \ref{fig:8}) after crossing the saddle point of the liquid nucleus and the free energy barrier in  Eq.~(\ref{eq:2-3x}) appears to be zero ($\Delta G_{\rm SL}=0$), there will be another free energy barrier that comes from the preexponential factor $A_{\rm SL}$ in Eq.~(\ref{eq:2-3x}).  Therefore the vapor to solid nucleation rate $J_{\rm SV}$ will be approximately given by Eq.~(\ref{eq:2-3xx}) and will be characterized by two nucleation rates $J_{\rm LV}$ and $J_{\rm SL}$.  Our simple capillarity theory cannot include such a kinetic effect as ours is based on the quasi-equilibrium thermodynamics and cannot include kinetic effect.
 
\section{\label{sec:sec4}Conclusion}
We have used the classical capillarity approximation to study the whole process of nucleation when an intermediate metastable phase is involved.  By following the minimum-free-energy path in the free-energy landscape of nucleation, we studied not only the critical nucleus at the saddle point of the energy landscape but also the whole process of nucleation starting from the initial embryo.  We found that the critical nucleus can be solid, composite, or liquid depending both on the relative supersaturation of the metastable liquid phase and on the wetting properties of the liquid and solid phases.  The compositions of solid and liquid in the composite nucleus depends strongly on the relative supersaturation of the intermediate liquid phase.  Although the free-energy barrier of the critical nucleus at the saddle point can also be studied using density functional theory, a comparison of the free energy only at the critical point does not reveal much about the nucleation process when an intermediate metastable phase is involved.  

Since we studied nucleation and not growth, the appearance of a macroscopic metastable phase during growth is outside the scope of this work.  For such a problem, various variants of the phase-field model~\cite{Bechhoefer1991,Celestini1994,Evans1997a,Iwamatsu2010} will be useful. Finally, we stress that our simple model based on the capillarity approximation predicted that the intermediate metastable state (liquid) can survive even when the metastable liquid phase is not critical~\cite{Tavassoli2002,Shiryayev2004}.  A divergent correlation length when the intermediate metastable liquid phase is critical~\cite{Tavassoli2002,Shiryayev2004} and the long-range intermolecular interaction~\cite{Bieker1998} will certainly affect the conclusions derived from our simplified capillarity theory of nucleation based on the short-range interaction. These issues are left for future investigations.

\begin{acknowledgments}
This work was supported by Grant-in-Aid for Scientific Research (C) 22540422 from Japan Society for the Promotion of Science (JSPS).   
\end{acknowledgments}

\end{document}